\documentstyle[preprint,aps]{revtex}
\begin{document}

\draft
%\twocolumn[

\title{ 
         The Finite-$U$ Impurity Anderson Model \\
         in the presence of an external magnetic field
           }

\author{Kicheon Kang and B. I. Min }

\address{ Department of Physics,
          Pohang University of Science and Technology\\
          Pohang 790-784, Korea }

\date{\today}

\maketitle

\begin{abstract}
We have investigated effects of an external magnetic field 
in the impurity Anderson model with a finite on-site Coulomb repulsion $U$.
Large $N_f$ expansion is employed in the slave boson representation,
by taking into account $f^0$, $f^1$, and $f^2$ subspaces.
To evaluate the vertex function for the ``empty state boson" self-energy,
we have devised two approximations which reduce much
computational efforts without losing general features of the model.
It is found that the Kondo temperature is reduced by the
presence of a magnetic field, and that
at low field and at low temperature,
the field dependence of both the Kondo temperature 
and the impurity magnetization exhibits a scaling behavior with 
high accuracy.
Further, some interesting features are found in the field dependence of the 
impurity magnetization at finite temperature, the physical implications 
of which are discussed in terms of the renormalized
Kondo temperature.
\end{abstract}
\pacs{PACS numbers: 71.10.+x, 71.27.+a, 71.28.+d, 75.20.Hr}

\section{Introduction}
Heavy fermion and valence fluctuation phenomena (see 
e.g., \cite{gunnar87,grewe91}) observed in Ce and Yb compounds 
have been often described by the impurity Anderson model~\cite{anderson61}. 
Due to a large on-site Coulomb repulsion $U$ between $f$-electrons,
these compounds are considered as typical examples of strongly 
correlated systems to which the conventional canonical perturbation 
technique cannot be applied.
A systematic large-$N_f$ ($N_f$ the spin and orbital degeneracy) 
expansion method has been used for the assumed
infinite-$U$ model to describe dynamical as well as thermodynamic 
properties\cite{gunnar83,bickers87}.
The infinite-$U$ treatment allows one to consider
only $f^0$ and $f^1$ subspaces of $f$-electrons, 
and thus makes the problem quite simple.
However, this scheme cannot be justified for realistic systems
because $U$ is not an infinity (5$\sim 6$\,eV typically).
Especially in Ce compounds, $f^2$ configurations are even energetically 
comparable to the $f^0$ configuration.

In the ``spin fluctuation" limit (or Kondo limit), 
the finite-$U$ effect is manifested in various
physical properties through the renormalized Kondo temperature $T_A$. 
The renormalization of $T_A$ arises from the change of the exchange 
coupling constant $J$ between conduction electron and $f$-electron spins
due to a finite $U$, in a sense of the Schrieffer-Wolff 
transformation~\cite{schrieffer66}.
While this effect was recognized in the exact Bethe 
ansatz approach~\cite{tsvelick83} and also in the numerical renormalization 
group approach~\cite{krishna80} for $N_f=2$, more appropriate
descriptions of the systems with $N_f>2$
were provided by the large-$N_f$ treatments including 
the $f^2$- subspace in the variational 
approach~\cite{gunnar87,gunnar85} and in the 
perturbational 
approach~\cite{sakai88,holm89,pruschke89,keiter90,schiller93,holm93}.
Using the generalized slave boson technique,
Schiller and Zevin~\cite{schiller93} have shown that the lowest
order perturbation expansion reproduces the exact $N_f\rightarrow\infty$
variational result at zero temperature ($T=0$) \cite{gunnar85}.

In this paper, 
we have investigated effects of an external magnetic field ($H$)
in the finite-$U$ impurity Anderson model,
employing the slave boson technique~\cite{coleman84} 
which is generalized to the finite-$U$ case \cite{holm89}.
To our knowledge, there has been no previous study of considering 
both the finite-$U$ and the magnetic field effect,
simultaneously.
We obtain the Kondo temperature and the impurity magnetization as
a function of $U$ and $H$. The impurity magnetization at a finite 
temperature is also determined.
In the finite-$U$ case, a set of coupled integral equations appears
even in the lowest order $1/N_f$ expansion.
In order to solve the coupled integral equations, we have exploited simplifying 
approximations which reduce substantial computational
efforts. This treatment allows one to study easily even 
anisotropic systems, such as those with an external magnetic field 
or with crystalline fields.

The paper is organized as follows. In section II, the large-$N_f$ 
slave boson theory is formulated for the finite-$U$ impurity Anderson
model. In section III, approximation schemes used 
in evaluating the integral equations are described,
and the $U$-dependence of the Kondo temperature is provided.
Section IV is devoted to studying the model in 
the presence of the external magnetic field both at $T=0$
and at a finite temperature. A special attention is paid to the field
dependence of the magnetization. Finally, conclusions are 
summarized in section V. 

\section{Large-$ N_{\lowercase{f}} $ Slave Boson Treatment}

We consider the finite-$U$ impurity Anderson Hamiltonian in a partial
wave representation:
\begin{equation}
{\cal H} = \sum_{k,m} \varepsilon_k\, c_{km}^{\dagger}c_{km} +
  \sum_m \varepsilon_{f_m}\, f_m^{\dagger}f_m
  + \sum_{k,m} V(k) \left( c_{km}^{\dagger}f_m + f_m^{\dagger}c_{km} \right)
  + U\sum_{m'>m} n_{m'}n_m,  \label{eq:hamil}
\end{equation}
where the indices $m,m'(=1,2,\cdots N_f)$ denote 
spin and orbital quantum numbers.
The operator $c_{km}^{\dagger}(c_{km})$ creates (annihilates) a conduction
electron in the state $(k,m)$ with an energy dispersion of
$\varepsilon_{\bf k} =\varepsilon_k$, $k\equiv |{\bf k}|$.
The $f$-electrons of the magnetic impurity have $m$-dependent 
energy $\varepsilon_{f_m}$.
The corresponding creation (annihilation) operator
and the number operator are $f_m^{\dagger}(f_m)$ and
$n_m=f_m^{\dagger}f_m$, respectively.
The matrix element
$V(k)$ represents the hoping of an electron between the
conduction band and the impurity
level. We assume that the hoping takes place only between the states
with the same $m$, and that the $V(k)$ is independent of $m$.
The Hamiltonian of Eq.(\ref{eq:hamil}) is studied in the subspace of at most
double occupancy, that is, in the subspace of $f^0,f^1$, and $f^2$ states.
This is a good starting point, since $U$ is large enough to rule out
configurations with more than double occupancies. 

Coleman~\cite{coleman84} has introduced a single slave boson $b$
to treat the infinite-$U$ model. Then standard many-body techniques can be
utilized in this approach without losing the strongly correlated properties
of the model. Coleman's method can be generalized 
to study the finite-$U$ model, by introducing a set of `heavy bosons'
$d^{\dagger}_{mm'}= -d^{\dagger}_{m'm}$ \cite{holm89,schiller93,holm93}.
The heavy bosons represent $\frac{N_f(N_f-1)}{ 2 }$ doubly occupied $f^2$ 
states, energies of which  are $E_{2mm'}=\varepsilon_{f_m}+
\varepsilon_{f_{m'}}+U$.
In this scheme, the Hamiltonian becomes
\begin{eqnarray}
 {\cal H } &=& \sum_{k,m} \varepsilon_k\, c_{km}^{\dagger}c_{km} +
   \sum_m \varepsilon_{f_m}\, f_m^{\dagger}f_m
   + \sum_{m>m'} E_{2mm'}d^{\dagger}_{mm'}d_{mm'}  \nonumber \\ 
   &+& \sum_{k,m} V(k) \left( c_{km}^{\dagger}b^{\dagger}f_m + \mbox{h.c.} 
                     \right)
   + \sum_{k,m\ne m'} V(k) \left( c_{km}^{\dagger}f^{\dagger}_{m'}d_{mm'}
   + \mbox{h.c.}     \right) .
\end{eqnarray}
This Hamiltonian commutes with the ``charge" operator $Q$ defined by
\begin{equation}
 Q\equiv b^{\dagger}b + \sum_m f_m^{\dagger}f_m +
 \sum_{m>m'}d^{\dagger}_{mm'}d_{mm'} .
\end{equation}
Physical quantities of the impurity Anderson model should be obtained 
with the constraint $Q=1$. One way to impose the constraint is to add a
`chemical potential' term $\lambda Q$ to ${\cal H}$, and to project it
onto the physical subspace $Q=1$ by taking
$\lambda\rightarrow\infty$ at the end of calculation ~\cite{coleman84}. 

With this prescription,
the partition function and other quantities are expressed in terms of
the spectral functions $\rho_{\alpha}(\omega)$, $\alpha=b,f_m,d_{mm'}$
corresponding to the $f^0,f^1,f^2$ states, respectively. 
The total partition function can be decoupled to a product $Z_cZ_f$,
where $Z_c$ is the partition function of the non-interacting conduction 
electrons and $Z_f$ is the impurity contribution given by
\begin{equation} 
 Z_f = \int_{-\infty}^{\infty} d\omega\,e^{-\beta\omega} \left(
  \rho_b(\omega)+\sum_m \rho_{f_m}(\omega)+\sum_{m>m'}\rho_{d_{mm'}}(\omega)  
                                      \right) . \label{eq:partition}
\end{equation}
Each of the spectral function is related to the imaginary part
of the corresponding retarded Green's function $G_{\alpha}(\omega)$ :
\begin{equation}
 \rho_{\alpha}(\omega) = -\frac{1}{\pi} Im\,G_{\alpha}(\omega) .
% \;\;\; \delta\mbox{ a positive infinitesimal number }
\end{equation}
The Green's functions are written in terms of their
self-energies $\Pi$, $\Sigma_{f_m}$, $D_{mm'}$ :
\begin{mathletters}
\label{eq:green}  
\begin{eqnarray}
 G_b(\omega) &=& \left( \omega-\Pi(\omega)+i\delta \right)^{-1}
      \\
 G_{f_m}(\omega) &=& \left( \omega-\varepsilon_{f_m}-
   \Sigma_{f_m}(\omega)+i\delta \right)^{-1}   \\
 G_{d_{mm'}}(\omega) &=& \left( \omega-E_{2mm'}-
   D_{mm'}(\omega)+i\delta \right)^{-1} ,
\end{eqnarray}
\end{mathletters}
with a positive infinitesimal number $\delta$.
The self-energies can be evaluated by using the ``Non-Crossing 
Approximation (NCA)" which is generalized for 
the finite-$U$~\cite{sakai88,holm89,pruschke89,keiter90,schiller93}.
Recall that the NCA has been very successful for the infinite-$U$ 
model \cite{bickers87,coleman84,kuramoto83}.
The ``generalized NCA" scheme leads to  coupled integral equations 
containing complicated vertex corrections. For this reason,
it is practically impossible to solve the equations exactly, and so
some simplifying approximations are necessary to resolve the problem.

Instead of using the generalized NCA, we adopt here a simple
picture of considering only the lowest
order diagrams in $1/N_f$, that is $(1/N_f)^0$.
In fact, crucial finite-$U$ effects are incorporated already 
in the lowest order diagrams, as shown below. 
Figure \ref{fig:diagrams} shows the lowest order diagrams for
the self-energies of the empty boson and the pseudo-fermion $f$.
The self-energy of the heavy $d$-boson 
has no contribution in the lowest order. 
Diagrams in Fig.\ref{fig:diagrams} are calculated 
by using the standard Feynman rules with the projection 
procedure mentioned above: 
\begin{mathletters}
\begin{equation}
 \Pi(\omega) = \frac{\Delta}{\pi}\sum_m \int_{-B}^{B}d\varepsilon\,
  f(\varepsilon) G_{f_m}(\omega+\varepsilon) \Gamma_m(\omega ; \varepsilon)
    \label{eq:pi}
\end{equation}
with the vertex function,
\begin{equation}
 \Gamma_m(\omega ; \varepsilon) = 1 + \frac{\Delta}{\pi}\sum_{m'(\ne m)}
  \int_{-B}^{B}d\varepsilon' f(\varepsilon') G_{f_{m'}}(\omega+\varepsilon') 
   G_{d_{mm'}}(\omega+\varepsilon+\varepsilon') 
   \Gamma_{m'}(\omega ; \varepsilon'),
      \label{eq:vertex}
\end{equation}
and
\begin{eqnarray}
 \Sigma_{f_m}(\omega) &=& \frac{\Delta}{\pi}\sum_{m'(\ne m)}
  \int_{-B}^{B} d\varepsilon' f(\varepsilon) G_{d_{mm'}}(\omega+\varepsilon)
     \label{eq:sigma_f} ,   \\
 D_{mm'}(\omega) &=& 0  . \label{eq:self_d}
\end{eqnarray}
\end{mathletters}
Here $f(\varepsilon)$ is the Fermi-Dirac distribution function, and
 $\Delta\equiv \pi\rho(\varepsilon)V^2(\varepsilon)$ is 
the hopping rate of an electron
between conduction band and impurity state in a magnetic channel. 
Here $\rho(\varepsilon)$ is the density of states of the 
conduction band.  The hopping rate is
assumed to be a constant $\Delta$ for $-B<\varepsilon<B$ and 0 otherwise. 
Eqs.(\ref{eq:pi}) - (\ref{eq:self_d}) with Eq.(\ref{eq:green})
complete the theory of the lowest order in $1/N_f$. 
The present treatment reproduces at $T=0$
the exact $N_f\rightarrow\infty$ results of 
the variational approach~\cite{gunnar85,schiller93}.

In the spin fluctuation limit, 
where $f^0$ and $f^2$ configurations
are energetically unfavorable relative to $f^1$,  most 
important properties of the Anderson model are determined by
the $b$-boson self-energy $\Pi$. The $\Pi$ contains
diagrams representing successive ``spin-flip" scatterings, 
which consist of two kinds of processes
as shown in Fig.\ref{fig:spinflip}.
These are the processes which give rise to the coupling constant 
$J\sim \frac{|V|^2}{\varepsilon_f} - \frac{|V|^2}{(\varepsilon_f+U)}$ between
conduction electron spin and the impurity spin, in
the Schrieffer-Wolff transformation~\cite{pruschke89}.
While the process (a) in the figure is present 
in the infinite-$U$ model having an $f^0$ intermediate
state, the process (b) which contains the $d$-boson line as 
an intermediate state does not exist in the infinite-$U$
model. The process (b), which comes out from the vertex 
function $\Gamma_m$ in the empty boson self-energy,
leads to a renormalization of the Kondo temperature 
$T_A$ through the modification of the coupling constant $J(U)$ and
$T_A(U)\sim B\exp{ \left[ 1/N_fJ(U)\rho(0) \right] }$. 

\section{Approximations and the Kondo temperature}
 
Thermodynamic properties of the finite-$U$
impurity Anderson model in the restricted subspaces $f^0, f^1$, and $f^2$ 
are determined entirely by the spectral functions $\rho_{\alpha}(\omega)$.
The ground state energy $E_0$ is obtained from
the lowest pole of $\rho_b(\omega)$ at $T=0$, 
which corresponds to a non-magnetic ground state. 
Using Eq.(\ref{eq:pi}),
one gets the following equation for $E_0$ :
\begin{equation}
 E_0 = \Pi(E_0),   \label{eq:gs}
\end{equation}
with $\Pi$ calculated at $T=0$, where $E_0$ is real.
Similarly, the lowest poles of $\rho_{f_m}(\omega)$ yield 
magnetic excited energies $\tilde{\varepsilon}_{f_m}$ satisfying 
\begin{equation}
 \tilde{\varepsilon}_{f_m} - {\varepsilon}_{f_m} 
   -\Sigma_{f_m}(\tilde{\varepsilon}_{f_m}) = 0  .
\end{equation}
$\Sigma_{f_m}(\tilde{\varepsilon}_{f_m})$ has no imaginary part
in the temperature range of interest, $T\lesssim T_A$. 
The ``Kondo temperature" $T_A$, which is defined by the difference 
between the ground state energy $E_0$ and the lowest excited state energy 
$\min{ \{\tilde{\varepsilon}_{f_m}\} }$, 
characterizes the low temperature and low energy properties of the model.
$E_0$ and $\tilde{\varepsilon}_{f_m}$ calculated in this fashion 
reproduce the exact $N_f\rightarrow\infty$ 
ground state and the excited state energy, respectively, obtained by the 
variational approach~\cite{gunnar85,schiller93}.

To get $E_0$, one should solve numerically the set of coupled integral 
equations involving the vertex function $\Gamma_m$.
Note that in the infinite-$U$ case, $\Gamma_m=1$. 
To solve the equation, we have devised two simple 
approximations, which can be readily applied
to anisotropic systems without much computational efforts.
In $\Gamma_m$, there are two independent energy 
variables, $\omega$ and $\varepsilon$. 
While $\Gamma_m$ has a large dependence on $\omega$ near $\omega=E_0$,
due to $G_{f_{m'}}(\omega+\varepsilon')$ in the integrand of 
the Eq.(\ref{eq:vertex}), 
it has relatively a weak dependence on $\varepsilon$. The $\varepsilon$
dependence of $\Gamma_m$ exists only
in $G_{d_{mm'}}(\omega+\varepsilon+\varepsilon') = \left(
\omega+\varepsilon+\varepsilon'-E_{2mm'} \right)^{-1}$. This dependence would
be weaker for larger $U$.
Thus we replace $\Gamma_{m'}(\omega;\varepsilon')$ in the integrand by 
$\Gamma_{m'}(\omega;\varepsilon)$ to avoid solving the coupled 
integral equations.
%
% inserted....
Actually, this corresponds to neglecting some energy conservation of 
diagrams for vertex function in Fig.1. However, owing to the weak dependence
on $\varepsilon$ of the Eq.(\ref{eq:vertex}), 
this approximation turns out to be quite good, as will be shown below. 
The resulting equation reads
\begin{mathletters}
\begin{equation}
 \Gamma_m(\omega ; \varepsilon) = 1 + \frac{\Delta}{\pi}\sum_{m'(\ne m)}
  \Lambda_{mm'}(\omega ; \varepsilon) \Gamma_{m'}(\omega ; \varepsilon)  ,
     \label{eq:vertex2}
\end{equation}
where
\begin{equation}
 \Lambda_{mm'}(\omega ; \varepsilon) = \int_{-B}^{B}d\varepsilon' 
   f(\varepsilon') G_{f_{m'}}(\omega+\varepsilon')
   G_{d_{mm'}}(\omega+\varepsilon+\varepsilon')  .
      \label{eq:Lambda}
\end{equation}
\end{mathletters}
Now Eq.(\ref{eq:vertex2}) is merely a linear algebraic equation 
of $\Gamma_m$'s for given energy variables $\omega,\varepsilon$. 
This is our first approximation (I).

The equation can be further simplified by representing the pseudo-fermion 
Green's function as
\begin{mathletters}
\begin{equation}
 G_{f_m}(\omega) = \frac{ z_{f_m} }{ \omega-\tilde{\varepsilon}_{f_m}+i\delta }
   \;\;\; ,    \label{eq:Gf}
\end{equation}
where $z_{f_m}$ is the renormalization coefficient at $\omega=
\tilde{\varepsilon}_{f_m}$ defined by
\begin{equation}
 z_{f_m} = \left. \left( 1 - \frac{ \partial }{ \partial\omega }\mbox{Re\,}
                 \Sigma_{f_m}(\omega)
                 \right)^{-1} 
         \right|_{ \omega=\tilde{\varepsilon}_{f_m} } .
\end{equation}
\end{mathletters}
This representation is quite reasonable, since the
incoherent background of $G_{f_m}$ for large $U$ is located at
very high energy ($\omega>E_{2mm'}$) as compared to $\tilde{\varepsilon}_{f_m}$,
%
%  inserted....
and the weight of the incoherent part ($1-z_{f_m}$) would be very small
(note that it becomes zero for infinite $U$). Thus representing $G_{f_m}$ by
Eq.(\ref{eq:Gf})
makes no practical difference in evaluating Eq.(\ref{eq:Lambda}) for
large $U$.  Then $\Lambda_{mm'}$ in Eq.(\ref{eq:Lambda}) 
can be obtained analytically at $T=0$,
\begin{equation}
 \Lambda_{mm'}(\omega ; \varepsilon)
    = \frac{ z_{f_{m'}} }{ E_{2mm'}-\tilde{\varepsilon}_{f_m}-\varepsilon } 
    \log \left ({ \frac{ \tilde{\varepsilon}_{f_m}-\omega+B }
    { \tilde{\varepsilon}_{f_m}
   -\omega } \frac{ E_{2mm'}-\omega-\varepsilon }
    { E_{2mm'}-\omega-\varepsilon+B } }\right ) .
\end{equation}
This is our second approximation (II).

%We apply our approximation schemes to discuss
%thermodynamics of the fully degenerate systems ($\tilde{\varepsilon}_{f_m}
%=\tilde{\varepsilon}_{f}$). 
The present approximation schemes 
yield fairly good results of the ground state energy $E_0$ and the 
Kondo temperature $T_A$ for $U\gtrsim 5$\,eV.
As shown in Fig.\ref{fig:kondo}, in the large $U$ limit ($U\gg B$), 
the Kondo temperature as a function of $1/U$ increases or decreases
depending on the parameters, $\varepsilon_f$, $\tilde{\Delta}=N_f\Delta$,
and $B$~\cite{gunnar85}. 
Results obtained by using both approximations (I and II) are 
close to those obtained by using the approximation (I) only.
Furthermore, both results are in good agreement with those of 
the exact $N_f\rightarrow\infty$ variational approach.
This demonstration indicates that the approximations (I) and (II)
we employed are very reasonable
%
% ..inserted...
at least relatively large $U$. Deviations from the variational calculations
for smaller $U$ appear because the approximation schemes made above become
less exact for smaller value of $U$, as pointed out above.    
We used in Fig.\ref{fig:kondo} quite a large value of the half-bandwidth
$B=24$\,eV.  One can conjecture that the 
approximation (I) works better for a realistic smaller bandwidth, 
{\it e.g.} $2B<O(10\mbox{eV})$, because, for smaller $B$,
the $\varepsilon'$-dependence of the vertex function 
$\Gamma_{m'}(\omega,\varepsilon')$ in the integrand of
Eq.(\ref{eq:vertex}) would be weaker.

\section{Scalings in the presence of the external magnetic field}
With the approximation schemes described in the previous section,
we investigate effects of the external magnetic field in
the finite-$U$ impurity Anderson model.
By applying the external field $\vec{H}=H\hat{z}$ to a degenerate system
($\varepsilon_{f_m}=\varepsilon_f$), 
Zeeman splitting occurs in the localized
$f$-levels; ${\varepsilon}_{f_m}={\varepsilon}_{f}+g\mu_B Hm$,
$m=-j,-j+1,\cdots,j$, $(2j+1=N_f)$. 
Conduction electron polarization is neglected, 
since its contribution is perturbatively small~\cite{lowenstein84}.

The Kondo temperature in the presence of $H$ is also obtained from 
the energy difference between the ground state and the lowest excited state 
energy. Figure \ref{fig:kondoH} plots the magnetic field dependence of 
$T_A$.  $T_A$'s are normalized by their zero-field values as
given in Table I.
It is seen that $T_A$ decreases with increasing $H$.
A reduction of the Kondo temperature by the applied field 
originates from the reduction of the ground state 
binding energy. The applied field lifts up the degeneracy of the $f$-level
to decrease the effective degeneracy.
Hence the binding energy will be reduced exponentially, recalling that
$T_A\sim B\exp{ \frac{\pi\varepsilon_f}{N_f\Delta} }$ in the Kondo limit
of the infinite-$U$ model.
Notable in this figure is that $T_A(H)/T_A(0)$ exhibits almost the 
same functional form against $g\mu_B H/k_BT_A(0)$
regardless of the size of $U$, implying that the scaling behavior
holds with high accuracy in the low field region. 
In contrast, $T_A(H)/T_A(0)$ has a different scaling behavior for 
different $N_f$. 
% The larger the $N_f$, the smaller the Kondo temperature.
In the infinite-$U$ limit, $T_A(H)/T_A(0)$ can be expressed as
(see Appendix),
\begin{equation}
 \frac{ T_A(H) }{ T_A(0) } =   \frac{ 1 }{  \prod_{m=1}^{N_f-1} 
  \left[
    \left( 1+m\frac{ g\mu_BH/T_A(0) }{ T_A(H)/T_A(0) } \right)
  \right]^{1/N_f}                        }
   \exp{ \left[
    \frac{ \pi }{ \tilde{\Delta} } \left( E_0(H)-E_0(0) \right)
                         \right] }      . 
   \label{eq:kondoH}
\end{equation}
The solution of Eq.(\ref{eq:kondoH}) can be regarded as an $N_f$-dependent
scaling function characterizing the low $T$ and low $H$ properties of the 
model. Using the Kondo temperature $T_A$ rescaled by 
$U$, the scaling of Eq.(\ref{eq:kondoH}) will be also valid in the 
finite-$U$ case.

The impurity contribution to the magnetization $M^{imp}$ of the
system can be obtained from the relation
\begin{equation}
 M^{imp} = \frac{1}{\beta} \frac{\partial}{\partial H} \log{Z_f}.
\end{equation}
The ground state is no longer non-magnetic in the presence of the
magnetic field, since the field polarizes both the impurity electrons 
and the conduction electrons.
In Fig.\ref{fig:mag0}, the ground state impurity magnetization 
$M^{imp}(H)$ is plotted as a function of $g\mu_BH/k_BT_A(0)$.
The magnetization in this figure is normalized by
its saturation value $M^{imp}_{sat}=jg\mu_B$.
The magnetization at low field also shows a scaling behavior 
almost perfectly, as in the case of $T_A(H)$. 
The present results of $M^{imp}(H)$ are compared with the 
Bethe ansatz results for the Coqblin-Schrieffer model~\cite{hewson83}. 
In the Bethe ansatz results, appears the low temperature scale $T_L$
which is related to the zero temperature susceptibility 
$\chi_0=(g\mu_B)^2j(j+1)/3T_L$. On the other hand, 
the susceptibility in our treatment is given by
$\chi_0=n_f\frac{ (g\mu_B)^2j(j+1) }{ 3T_A }$ for $U=\infty$.
By considering the relationship between two energy scales,
it is possible to compare our results with the Bethe ansatz results.
As is seen in Fig.\ref{fig:mag0}, the present results of $M^{imp}(H)$
agree quite well with the Bethe ansatz results
in the low field region,
even though they are a bit overestimated in the high field region. 
The deviation from the exact results at high field indicates 
that the $1/N_f$ expansion may not work well in the high field region.
As mentioned previously, the applied field splits the $f$-level and 
reduces the effective degeneracy, and so the lowest order $1/N_f$ expansion 
becomes invalid in the high field region. 
It is noteworthy that the overestimation of $M^{imp}(H)$ observed in 
Fig.\ref{fig:mag0} is also found in the infinite-$U$ NCA scheme for large
$H/T$~\cite{cox87}.

An interesting feature is found in the low field
magnetization. The inset of Fig.\ref{fig:mag0}
indicates that the low field magnetization increases
more rapidly than linearly in $H$.
This deviation from the linearity, so called
the ``superlinear" behavior, has been
reported by  Hewson {\it et al}~\cite{hewson83},
based on the Bethe ansatz solution of the Coqblin-Schrieffer model.
They found that the superlinearity occurs for $N_f>3$, and
that the experimental results of YbCuAl are well 
described by their $N_f=8$ model.
The superlinearity is also revealed by 
the finite temperature NCA scheme~\cite{cox87} as well as
by the mean field approximation~\cite{newns84} in the infinite-$U$,
large-$N_f$ treatments of the Anderson model.
The superlinear behavior has its origin in the 
fact that the Kondo resonance, which is located near $T_A$ above the 
Fermi level, becomes narrower as $N_f$ increases.
That is, the Kondo resonance is sharply defined for large $N_f$,
implying a strong screening effect.
As a consequence, the spin polarization does not begin to dominate,
until the magnetic field becomes comparable to $T_A$.
As the Zeeman energy is comparable to $k_BT_A$, the
magnetic field overcomes the screening by conduction electrons,
and the spin polarization dominates the system.
It is quite natural that the
superlinearity is present in our finite-$U$ model, in view of 
the almost perfect scaling of $M^{imp}(H)$ in the low field regime. 
The superlinearity persists at low
$T/T_A$, but disappears at higher temperature because of
suppression of the Kondo resonance.

At finite temperature, the situation is more intriguing.
For a given low $T$,
$M^{imp}(H)$ shows a reupturn at a certain field $H^*$
(denoted by arrows in Fig.\ref{fig:magT}),
apart from the superlinearity at lower field.
Making more investigation into this anomaly, one could find that the Kondo 
temperature at the reupturn point is nearly equal to a given
temperature of the system, that is, $T_A(H^*)\simeq T$. 
This phenomenon could be understood as follows. For
a given low $T$, $M^{imp}(H)$ has a superlinearity at low $H$, but
diminishes its increasing rate for intermediate values of $H$. 
As $H$ increases further, $M^{imp}(H)$ exhibits a reupturn at $H=H^*$
which fulfills $T_A(H^*)\simeq T$. 
It takes place because the population in the lowest 
excited magnetic state increases abruptly at the point 
satisfying $T_A(H^*)\simeq T$, producing observed reupturns at $H^*$.
It is shown in Fig.\ref{fig:magT}  that the reupturn 
smears out at higher temperature, since the Kondo resonance
is suppressed at high temperature. Another thing to note is that
the scaling of $M^{imp}(H)$ at finite $T$ is not so good 
as that at $T=0$.

Figure \ref{fig:magU} present $M^{imp}(H)$ for $U=5$\,eV 
at various temperatures.  Non-monotonicity as a function of $T$
is apparent in the low field magnetization.
With increasing the temperature, $M^{imp}(H)$
for a given small $H$, which is proportional to magnetic susceptibility,
increases first at low $T$, but after reaching a maximum at 
an intermediate $T^*$, $M^{imp}(H)$ decreases at higher $T$
 (see the inset of the Fig.\ref{fig:magU}). 
%In Fig.\ref{fig:magU}, we have $T^*/T_A\simeq 0.22$. 
This non-monotonic behavior in the temperature dependence of
the magnetic susceptibility is known to be 
more pronounced for a system with larger $N_f$~\cite{rajan83},
and is considered to have the same origin as that of the
superlinearity in $M^{imp}(H)$~\cite{cox87}.

\section{Conclusion}
We have studied the finite-$U$ impurity Anderson model in the presence of an 
external magnetic field. Large-$N_f$ expansion considering $f^0$, $f^1$, and
$f^2$ subspaces in the slave boson representation is employed to take into
account finite-$U$ effects. The lowest order treatment gives rise to a 
complicated vertex function in the empty boson self-energy 
and a renormalization of $f$-level energies.
% It gives rise to the same equations as those of the exact 
%$N_f\rightarrow\infty$ variational approach at the zero 
%temperature~\cite{gunnar85,schiller93}. 
We have devised two simple approximation schemes to evaluate the vertex 
function. One consists in transforming the integral equation into an 
algebraic equation by assuming that the vertex function has
a weak dependence on one of two energy variables.
The other is to simplify the pseudo-fermion 
Green's function by neglecting the incoherent part.
These approximations reduce computational efforts substantially, 
and are shown to yield fairly good results for $U\gtrsim 5$\,eV. 

This treatment has been applied to the system with an external magnetic
field. It is found that the Kondo temperature is reduced by the
magnetic field. Scaling behaviors in the field dependence of
the Kondo temperature and the impurity magnetization
are found to hold almost perfectly at low field and low temperature. 
This implies that the main effect of the finite-$U$ in the presence
of the magnetic field is manifested through the renormalization of 
the Kondo temperature.
Some intriguing features are found in the
field dependence of the magnetization, such as superlinearity at low 
field and reupturns at a higher field $H^*$ for a given low
temperature, which are expected to occur from a competition between
the singlet binding energy and the Zeeman energy gain of electrons.
%due to magnetic moment of $f$-electron.

%
\section*{acknowledgments}
This work was supported by the POSTECH-BSRI program of the
Korean Ministry of Education and in part by the Korea
Science Engineering Foundation through the SRC program of
SNU-CTP.

%
%%-----------Appendix-----------------------------
%
\appendix
\section*{}
Let's consider the Kondo temperature in the presence of the applied 
magnetic field for $U=\infty$ case.
Degenerate $f$-levels split by the magnetic field.
The energy levels can be expressed in terms of the 
Kondo temperature, as defined in section III :
\begin{equation}
 \varepsilon_{f_m} = E_0(H)+T_A(H)+mg\mu_B H, \label{eq:split}
\end{equation}
where $m=0,1,2,\cdots,N_f-1$. In the $U\rightarrow\infty$ limit,
Eq.(\ref{eq:gs}) reduces to
\begin{equation}
 E_0(H) = \frac{\Delta}{\pi} \sum_m\int_{-B}^{0} 
    \frac{ d\varepsilon }{ E_0(H)-\varepsilon
     -\varepsilon_{f_m} }   \label{eq:gs2}
\end{equation}
at $T=0$.
Analytical evaluation of Eq.(\ref{eq:gs2}) with Eq.(\ref{eq:split})
yields
\begin{equation}
 E_0(H) = \frac{ \tilde{\Delta} }{ \pi } \log{ \frac{T_A}{B} 
  \left\{ \prod_{m=0}^{N_f-1}
     \left( 1+m\frac{g\mu_B H}{T_A(H)} \right) \right\}^{1/N_f} 
                                           } .  \label{eq:a3}
\end{equation}
In the derivation, we used the condition, $B\gg N_f g\mu_B H$ and $B\gg T_A$.
Eq.(\ref{eq:a3}) corresponds to Eq.(\ref{eq:kondoH}).

%%-----------References are given here------------
%

%% ----- Figure Captions ----------

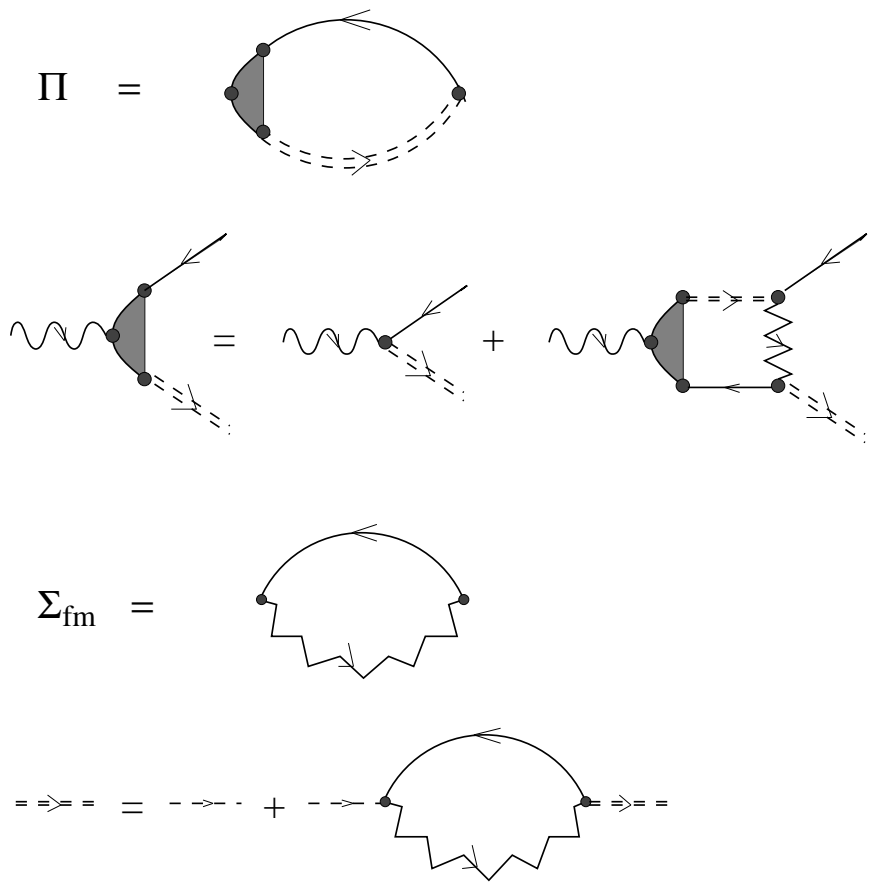
\begin{figure}
\caption{ Lowest order $\left( (1/N_f)^0 \right)$ diagrams of the
 empty boson and the pseudo-fermion self-energies in the finite-$U$
 Anderson model. The full, wiggly, dashed, and jagged lines stand for
 conduction electron, empty state boson, pseudo-fermion, and heavy boson
 propagators, respectively. Double dashed line denotes the renormalized
 pseudo-fermion propagator as represented by the diagram in the bottom. }
  \label{fig:diagrams}
\end{figure}
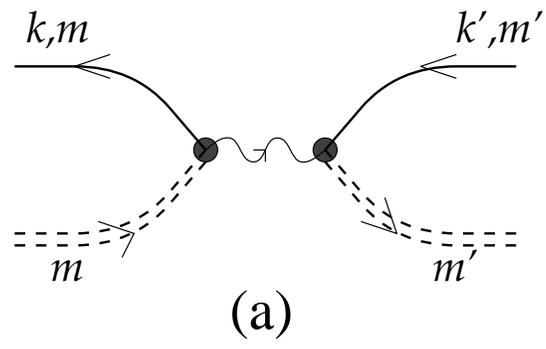
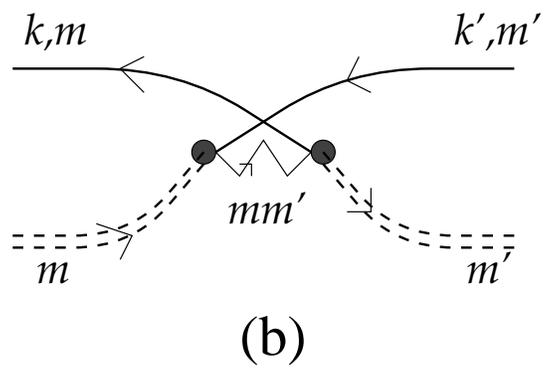
\begin{figure}
\caption{ Two elementary ``spin flip" scattering processes appearing
 in the empty boson self-energy $\Pi$, having an $f^0$-intermediate
 state (a) and an $f^2$-intermediate state (b). }
  \label{fig:spinflip}
\end{figure}
\begin{figure}
\caption{ Comparison of the present results for the Kondo 
 temperature with the exact $N_f\rightarrow\infty$ variational results
at $T=0$ (Ref. [9]).
 %by Gunnarsson and Sch\"{o}nhammer~\cite{gunnar85} (that is, the 
%by Gunnarsson and Sch\"{o}nhammer (Ref. [9]) 
%(that is, the 
% result from the Eq.(\ref{eq:pi}),(\ref{eq:vertex}) and
% (\ref{eq:gs}) at $T=0$. 
Dashed (solid) lines are the 
Kondo temperatures obtained by the approximation scheme (I)((I) and (II)),
and points correspond to the exact variational ones. Parameters used here are
$N_f=\infty$, $\tilde{\Delta}=N_f\Delta=0.75$\,eV, $B=24$\,eV with
 $\varepsilon_f=-2.5$\,eV in (a) and $\varepsilon_f=-1.5$\,eV in (b).  
Here degenerate $f$-level $\varepsilon_{f_m}=\varepsilon_f$ is assumed.}
  \label{fig:kondo}
\end{figure}
%
%\begin{figure}
%\caption{ (a) The ground state $f$-valency $n_f$ (solid line) and 
% (b) the Kondo temperature as a function of $1/U$. Parameters
% used are $N_f=6$, $\varepsilon_f=-2$\,eV, $\tilde{\Delta}=0.9$\,eV
% and $B=3$\,eV. The dashed line in (a) denotes the value of 
% $n_f/F(T_A)$ with $F(T_A)=\frac{ \tilde{\Delta}/\pi T_A }{ 
%  1+\tilde{\Delta}/\pi T_A }$. }
%   \label{fig:nf}
%\end{figure}
%
\begin{figure}
\caption{ Normalized Kondo temperature vs. normalized magnetic
 field strength for $N_f=4$ and $N_f=6$. 
The normalized Kondo temperatures are almost the same for different $U$'s. 
Other parameters used are $\varepsilon_f=-2$\,eV, 
$\tilde{\Delta}=0.9$\,eV, $B=3$\,eV. }
   \label{fig:kondoH}
\end{figure}
\begin{figure}
\caption{ Scaling behavior of the impurity magnetization which is
normalized by its saturation value $M^{imp}_{sat}=jg\mu_B$.
Bethe ansatz results of the Coqblin-Schrieffer 
model for $N_f=4(+)$ and $N_f=6(\Diamond)$ are also given, for the comparison. 
%Parameters used are the same as those in Fig.\ref{fig:kondoH} } 
%Parameters used are the same as those in Fig.4.}
The inset shows deviations of the low-field magnetizations from linearity for
 $N_f=4$ and $N_f=6$. Solid lines are the present results for $U=5$\,eV, and 
 dashed lines are extrapolated values from the initial slope, 
 $\left. \frac{ \partial m }{ \partial h } \right|_{h=0}h$. Other
%parameters are the same as those  in Fig.\ref{fig:kondoH}.  }
 parameters are the same as those  in Fig.4.}
   \label{fig:mag0}
%   \label{fig:superlinear}
\end{figure}
\begin{figure}
\caption{ Field dependence of the magnetization for $N_f=4$ at
 (a) $T/T_A=0.1$, (b) $T/T_A=0.3$, (c) $T/T_A=0.5$, and (d) $T/T_A=0.9$. 
%Other parameters are the same as those  in the Fig.\ref{fig:kondoH}.
Other parameters are the same as those in Fig.4.
Arrows in (a), (b), and (c) denote reupturns of the magnetization. 
The reupturn smears out at higher temperature as shown in (d).  }
   \label{fig:magT}
\end{figure}
\begin{figure}
\caption{ Field dependence of the magnetization for various 
 temperatures with $N_f=4$ and $U=5$\,eV. Other parameters used are
 the same as those in Fig.4. The inset shows the non-monotonic behavior
 of the magnetic susceptibility divided by its value of $T=0$, 
 $\chi(T)/\chi(0)$.}
%the same as those in Fig.\ref{fig:kondoH}.  }
   \label{fig:magU}
\end{figure}
%
%
%----------TABLES------------------------------

\begin{table}
 
\caption{ Kondo temperature $T_A$ at $H=0$ for $N_f=4$
 and $N_f=6$. Parameter sets are
%the same as those in Fig.\ref{fig:kondoH}. } 
the same as those in Fig.4.}
 
%\secdec 0.00
 \begin{tabular}{ccc}
  $U$\,(eV) & $T_A$\,(meV) $(N_f=4)$ & $T_A$\,(meV) $(N_f=6)$ \\
  \hline
  $\infty$ & 2.76 & 2.76 \\
   10  &  6.04 & 6.51 \\
   5   &  11.9 & 13.4
\end{tabular}
\end{table}


\begin{references}
%
\bibitem{gunnar87} See, e.g., O. Gunnarsson and K. Sch\"{o}nhammer, in
 {\em Handbook on the Physics and Chemistry of Rare Earths}, vol.10, ed. by
 K. A. Gschneidner Jr., L. Eyring and S. H\"{u}fner (North-Holland 1987)
 pp 103 - 163; and references therein.
\bibitem{grewe91} N. Grewe and F. Steglich, in
 {\em Handbook on the Physics and Chemistry of Rare Earths}, vol.10, ed. by
 K. A. Gschneidner Jr. and L. Eyring  (North-Holland 1991) pp 343 - 474;
and references therein.
\bibitem{anderson61} P. W. Anderson, Phys. Rev. {\bf 124}, 41 (1961).
\bibitem{gunnar83} O. Gunnarsson and K. Sch\"{o}nhammer, Phys. Rev. B
 {\bf 28}, 4315 (1983).
\bibitem{bickers87} N. E. Bickers, Review of Modern
 Physics {\bf 59}, 845 (1987); and references therein.
\bibitem{schrieffer66} J. R. Schrieffer and P. A. Wolff, Phys. Rev.
 {\bf 149}, 491 (1966); B. Coqblin and J. R. Schrieffer, Phys. Rev.
 {\bf 185}, 847 (1969).
\bibitem{tsvelick83} A. M. Tsvelick and P. B. Wiegman, Adv. Phys.
 {\bf 32}, 453-713 (1983).
\bibitem{krishna80} H. R. Krishna-murthy, J. W. Wilkins and K. G. Wilson,
 Phys. Rev. B {\bf 21}, 1003 (1980); {\bf 21}, 1044 (1980).
\bibitem{gunnar85} O. Gunnarsson and K. Sch\"{o}nhammer, Phys. Rev. B
 {\bf 31}, 4815 (1985).
\bibitem{sakai88} O. Sakai, M. Motizuki and T. Kasuya, in {\em Core-Level
 Spectroscopy in Condensed Systems}, Springer Series in Solid-State Sciences,
 ed., J. Kanamori and A. Kotani p.45 (Springer, Berlin, Heidelberg 1988).
\bibitem{holm89} J. Holm and K. Sch\"{o}nhammer, Solid State Commun.
 {\bf 69}, 969 (1989).
\bibitem{pruschke89} Th. Pruschke and N. Grewe, Z. Phys. B {\bf 74},
 439 (1989).
\bibitem{keiter90} H. Keiter and Q. Qin, Physica B {\bf 163}, 594 (1990);
 Q. Qin and H. Keiter, Z. Phys. B {\bf 84}, 89 (1991).
\bibitem{schiller93} A. Schiller and V. Zevin, Phys. Rev. B {\bf 47}, 9297
 (1993).
\bibitem{holm93} J. Holm, R. Kree and K. Sch\"{o}nhammer, Phys. Rev. B
 {\bf 48}, 5077 (1993).
\bibitem{coleman84} P. Coleman, Phys. Rev. B {\bf 29}, 3035 (1984).
\bibitem{kuramoto83} Y. Kuramoto, Z. Phys. B - Condensed Matter {\bf 53},
 37 (1983).
%\bibitem{rasul83} J. W. Rasul and A. C. Hewson, J. Phys. C: Solid State
% Phys. {\bf 16}, L933 (1983).
\bibitem{lowenstein84} J. H. Lowenstein, Phys. Rev. B {\bf 29}, 4120 (1984).
\bibitem{hewson83} A. C. Hewson, J. W. Rasul and D. M. Newns, Phys. Lett.
 {\bf 93A}, 311 (1983); A. C. Hewson and J. W. Rasul, J. Phys. C: Solid
 State Phys. {\bf 16}, 6799 (1983); A. C. Hewson, J. W. Rasul and D. M. Newns,
 Solid State Communications {\bf  47}, 59 (1983).
\bibitem{cox87} D. L. Cox, Phys. Rev. B {\bf 35}, 4561 (1987).
%\bibitem{mattens80} W. C. M. Mattens, Ph. D Thesis, Univ. of Amsterdam (1980).
\bibitem{newns84} D. M. Newns, N. Read and A. C. Hewson, in {\em Moment
 Formation in Solids}, ed. W. J. M. Buyers (Plenum, New York 1984), p.257;
 D. M. Newns and N. Read, Adv. in Phys. {\bf 36}, 799-849 (1987).
\bibitem{rajan83} V. T. Rajan, Phys. Rev. Lett. {\bf 51}, 308 (1983).
%
\end{references}
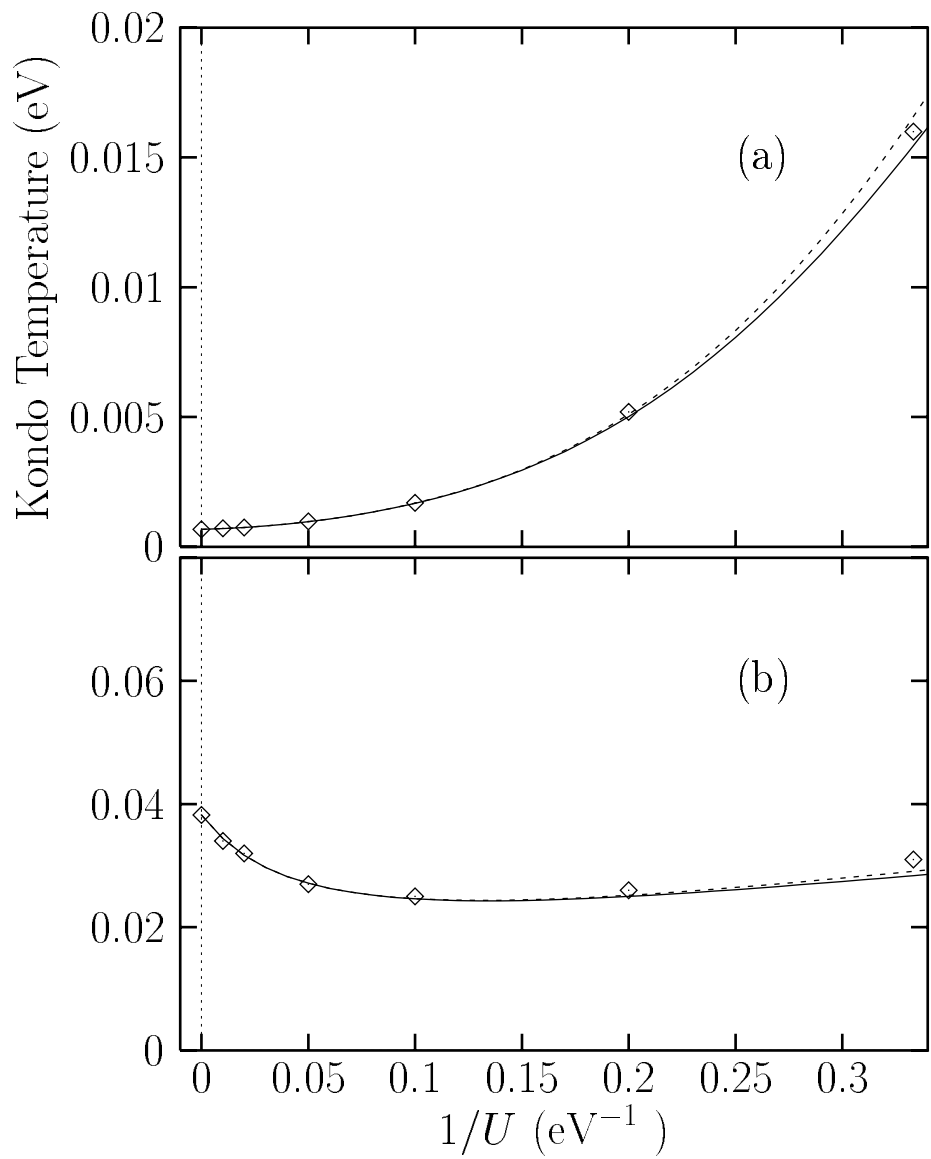
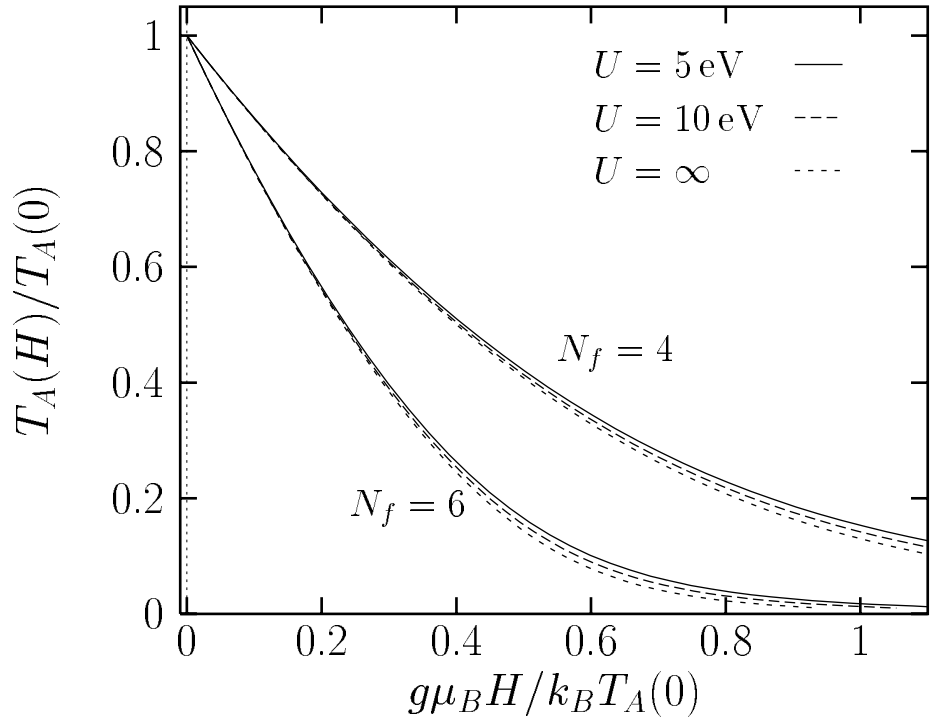
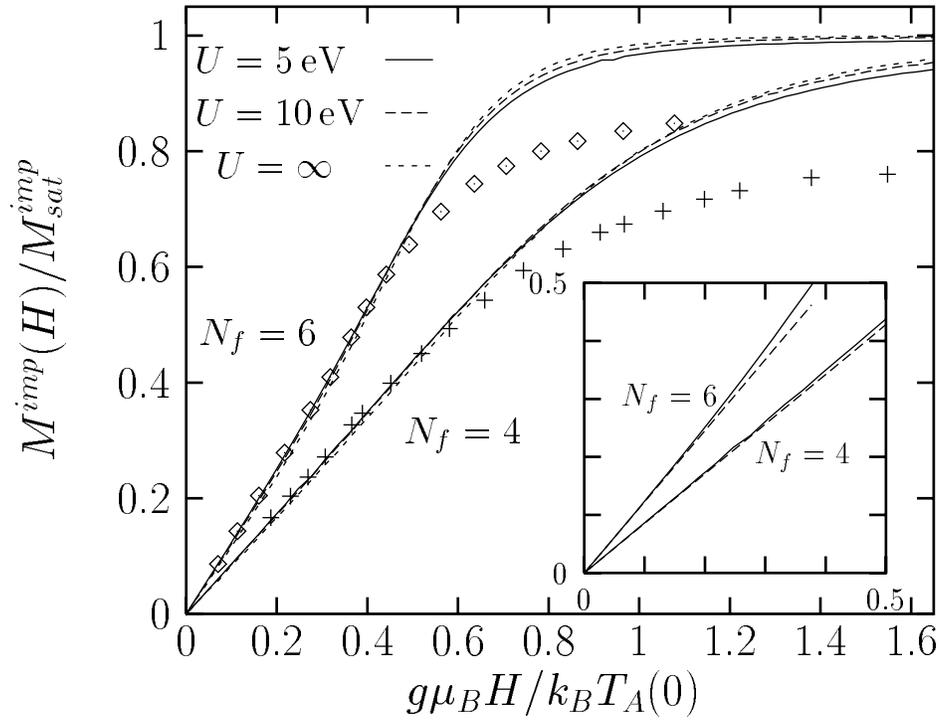
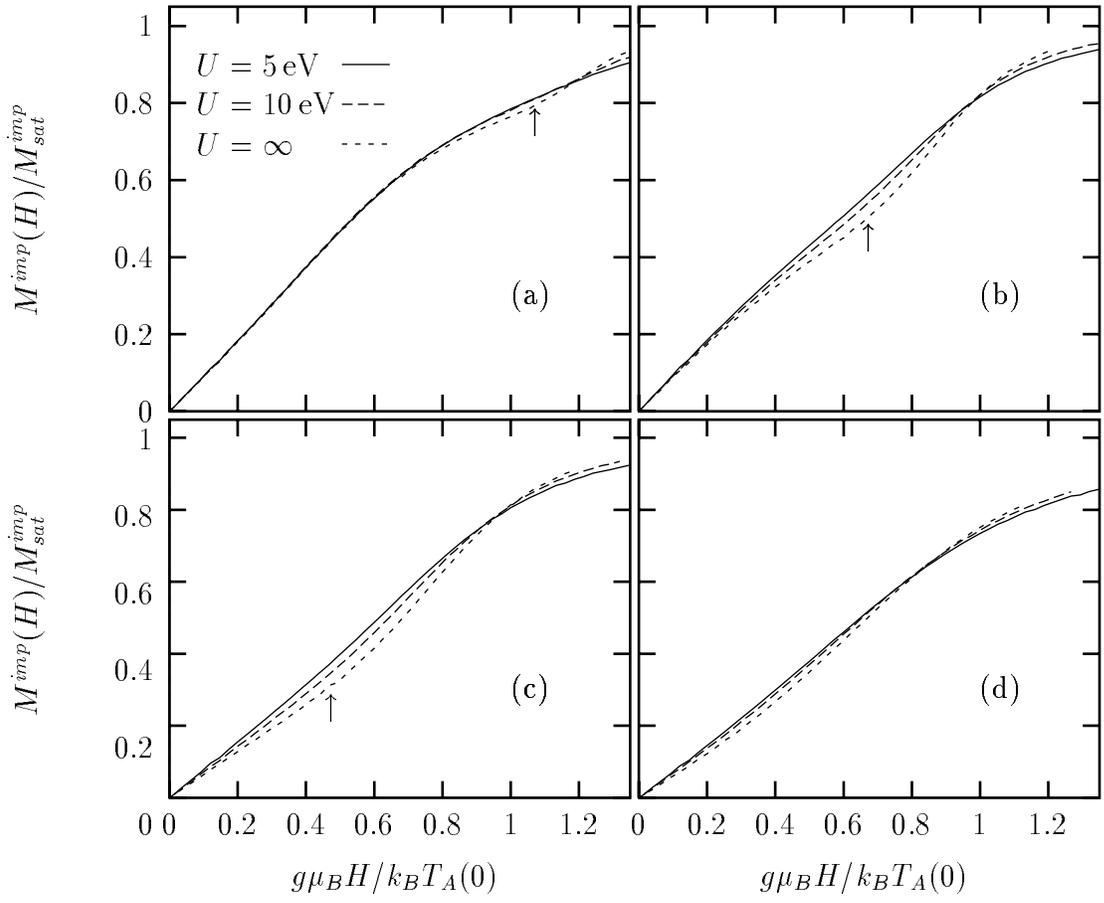
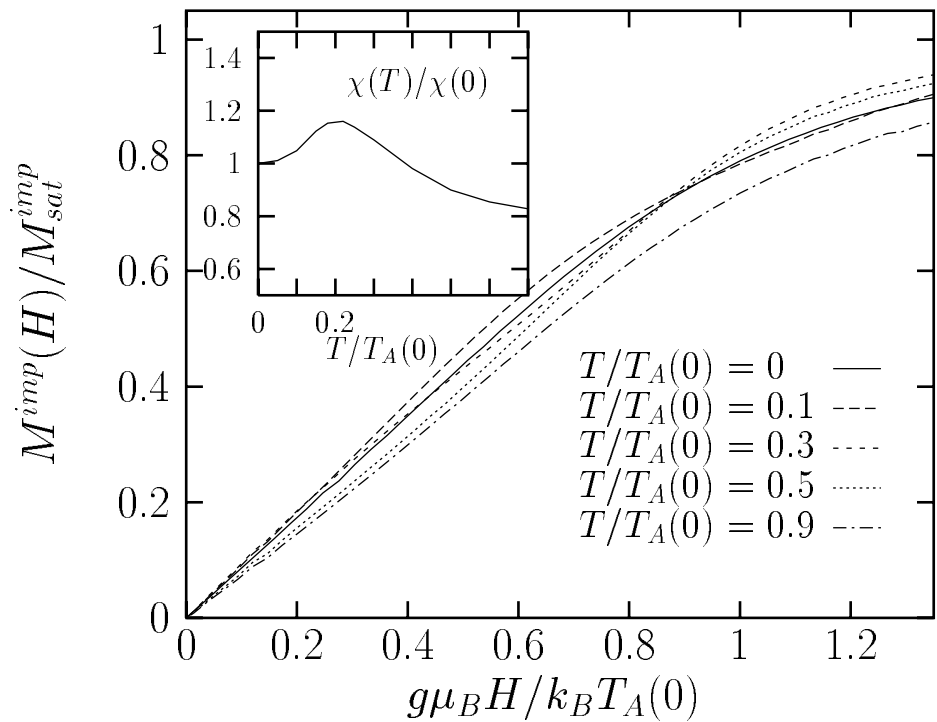
\end{document}